\UseRawInputEncoding
\documentclass[preprint,12pt,fleqn]{elsarticle}
\usepackage{amssymb,amsmath,amsthm}
\journal{and published in Annals of Physics, \emph{\textbf{430}, 168548-1/15, 2021}.                           \hskip200pt \tiny}
  
\newcommand{\bg} {\boldsymbol{g}}
\newcommand{\bgamma} {\boldsymbol{\gamma}}
\newcommand{\bchi} {\boldsymbol{\chi}}

\newcommand{\dpar}{\dot{\partial}}
\newcommand{\br} {\boldsymbol{r}}

\newcommand{\bL} {\boldsymbol{\mathcal{L}}}
\newcommand{\bmu} {\boldsymbol{\omega}}
\begin{document}

\begin{frontmatter}
\title{Electromagnetic and gravitational interactions\\ from Lagrangian mechanics}
\author{Paolo Maraner}
\ead{pmaraner@unibz.it}
\address{Faculty of Economics and Management, Free University of Bozen-Bolzano, Universit\"atsplatz 1 - Piazza Universit\`a 1, 39100~Bozen-Bolzano, Italy}
\begin{abstract} 
Background fields of electromagnetic and gravitational type emerge in the low kinetic energy limit of any regular Lagrangian system and, in particular, in the corresponding limit of any spacetime theory in which the free motion of test particles is described by an unspecified regular Lagrangian.  Electromagnetic and gravitational type interactions are therefore a universal feature of low kinetic energy Lagrangian systems. These background interactions can be consistently turned into dynamic Einstein-Maxwell fields by promoting the Lagrangian function  to a dynamic scalar field on the tangent bundle of the configuration space.   Accordingly, Einstein-Maxwell theory can be deduced from the assumption that the motion of elementary test particles in spacetime is described by Lagrangian mechanics. For higher kinetic energy-type values, identified with the square of the invariant mass of the particle, the Lagrangian induces higher rank interactions that seem however too weak to have been  detected in spacetime physics, but which might prove relevant at the Planck scale.
\end{abstract} 
\end{frontmatter}

\section{Introduction}
It has recently been shown~\cite{Maraner2019} that the trajectories of motion
of any regular autonomous Lagrangian system are the geodesics of an
energy-dependent Jacobi metric of Finsler type that in the limit of low kinetic
energies always reduces to a Randers metric.  Accordingly, the trajectories of
motion of any regular\footnote{Throughout this paper by  \textit{regular
Lagrangian} we mean a Lagrangian function $L(x,\dot{x})$ defined on the entire
tangent bundle, which is at least twice continuously differentiable in the base
coordinates $x$, analytic in the fiber coordinates $\dot{x}$ and such
that its Hessian in the fiber coordinates is everywhere non degenerate.}
Lagrangian system are, in this limit, indistinguishable from the paths in
configuration space of a representative point subject to interactions of the
electromagnetic and gravitational type. 
These interactions therefore have a universal character that is independent of the details of the Lagrangian system.
In particular, this type of interactions emerge in the corresponding limit of any spacetime theory in which the motion of test particles is described by an unspecified  regular Lagrangian. 
It is therefore natural to ask whether real electromagnetic and gravitational
fields also emerge in this way, thus representing a limit case of a more
general theory~\cite{LaemmerzahlandPerlick2018} based on Finsler~\cite{Rund1959,BCZ2012,Minguzzi2014} or Lagrange~\cite{Kern1974,Miron&Anastasiei1994} 
 geometry  rather than on the Riemannian one. To explore this possibility it is necessary to check whether the background fields induced by a regular autonomous Lagrangian can consistently be turned into dynamic fields or, equivalently, if the Lagrangian itself can be promoted to a dynamic field describing the geometry of the configuration space.

In this paper we will show that is possible to associate a non linear connection to a general regular Lagrangian in such a way that it can be promoted to a dynamic field with the induced interactions of electromagnetic and gravitational type that  evolve according to Einstein--Maxwell dynamics.    
As a guiding thread and since this is the case of greatest interest, we will consider a general Lagrangian describing the \textit{free} motion of elementary test particles in a four dimensional spacetime. However, beyond this possible application, our results extend to arbitrary dimensions and signatures and also to regular Lagrangians that include the scalar potential of an external force field. 
Our approach is close in spirit to the one recently advanced by Pfeifer and
Wohlfarth~\cite{PfeiferandWohlfarth2011,PfeiferandWohlfarth2012} for
Finsler gravity, but differs from previous attempts to build a Finsler theory
of the electromagnetic and gravitational fields for the choice of a general
regular free Lagrangian~\cite{Miron&Anastasiei1994}, rather than a Finsler
line element~\cite{Asanov1985,Javaloyes&Sanchez2014a,Javaloyes&Sanchez2014b,Javaloyes&Sanchez2020,BJS2020} or a Finsler Lagrangian~\cite{PfeiferandWohlfarth2012,Beem1970,Rutz1993,LPH2012,HassePerlick2019,HPV2019,Pfeifer2019}, as fundamental dynamic field. Correspondingly, we will not
have to deal with a single Finsler metric but with a one-parameter family of
Finsler metrics, one for each value of the energy-type first integral
associated to the autonomous Lagrangian, identified, when positive, with half
of the square of the invariant mass of the test particle. 
In the perturbative regime the Lagrangian also induces higher rank interactions
that produce Finsler modified dispersion relations of possible relevance in
quantum gravity  phenomenology~\cite{Amelino-Camelia2013}. In this respect our
approach is dual to
the construction of the spacetime Finsler structure starting from a non
homogeneous Hamiltonian through the Helmholtz constraint action, as recently
investigated in~\cite{Pfeifer2019,Girelli&al.2007,Ratzel&al.2011,Amelino-Camelia&al.2014,LetiziaandLiberati2017,LoboandPfeifer2020}.

In brief, our argument proceeds as follows. We will first show that for each energy-type value the trajectories in configuration space of any regular free Lagrangian system correspond to the geodesics of a Finsler metric, which in the limit of small energy-type values always reduces to a Randers metric. This step is crucial. On the one hand, it shows the geometrical character of the trajectories of the Lagrangian system. On the other hand, it gives us control of the small energy-type regime of the theory.   
In the application to spacetime physics this allows us to locate the evolution parameter that plays the role of   the proper time and to correspondingly identify the energy-type first integral, when positive, with half of the square of the invariant mass of the test particles.
For each energy-type value the geodesic equations will then be identified with the auto parallel equations of a non linear connection.
The margin of freedom inherent in this process will be discussed and fixed by requiring that the connection depends solely on the original Lagrangian and that the covariant derivative of the corresponding Hamiltonian is identically zero. 
In the application to spacetime physics this identifies the spacetime connection and curvature for each value of the invariant mass.
Finally, the Lagrangian will be promoted to a dynamic field in terms of a geometric  action integral on the configuration space tangent bundle in which a curvature scalar  is averaged over states with different energy-type values by a Gibbs measure. In the small energy-type limit this action always reduces to the Einstein--Maxwell action on the configuration space. In spacetime physics the Einstein--Maxwell theory can therefore be deduced from the basic assumption that the motion of elementary test particles is described by Lagrangian mechanics, supplemented by prescriptions for the non linear connection and for the action integral. These prescriptions are, on the other hand, completely general and purely geometric in nature.

To be specific, we will consider spacetime as a four-dimensional differentiable manifold $M$ locally parameterized  by the coordinates $x=\{x^i\}$, $i=0,1,2,3$. We will denote by $\dot{x}=\{\dot{x}^i\}$ the induced coordinates of a direction in the tangent space $T_xM$, so that the tangent bundle $TM$ is locally parameterized by the coordinates $(x,\dot{x})$.
We will also denote by $\partial_i=\frac{\partial\ }{\partial x^i}$ and $\dot{\partial}_i=\frac{\partial\ }{\partial \dot{x}^i}$ the partial derivatives with respect to $x^i$ and $\dot{x}^i$ respectively.
For greater clarity, with the sole exceptions of the Lagrangians  and the associated Hamiltonians, we will indicate in bold characters all quantities depending on both $x$ and $\dot{x}$. We will use the term `Finsler geometry' in its  broadest meaning of a geometry described by a differentiable line element $ds(x,dx)$ homogeneous of degree one in the infinitesimal displacements $dx$, 
\begin{equation}
ds(x,\lambda dx)=\lambda ds(x,dx)\hskip5pt\text{with}\hskip5pt\lambda>0,
\end{equation}
of which Riemannian geometry is a particular, but particularly significant, case. The hypothesis of reversibility, 
\begin{equation}
ds(x,-dx)=ds(x,dx),
\end{equation}
will also be considered. When not explicitly needed we will also omit the prefixes `pseudo' or `Lorentzian' to stress that the signature of the chosen spacetime structure is indefinite.

\section{Spacetime structures}
\label{Spacetime structures}
In general relativity the structure of spacetime is given in terms of a Lorentzian metric $g_{ij}(x)$, a non degenerate, symmetric, rank two, covariant tensor with signature $+,-,-,-$ (or $-,+,+,+$, depending on the convention). At each spacetime point $x$, a direction  $\dot{x}$ is classified as timelike, lightlike or spacelike according to whether the quadratic form $g_{ij}(x)\dot{x}^i\dot{x}^j$ is positive, zero or negative. Correspondingly, the affinely parameterized timelike and lightlike solutions of the Euler--Lagrange equations of the Lagrangian 
\begin{equation}
\label{LGR}
L_R(x,\dot{x})=\frac{1}{2}g_{ij}(x)\dot{x}^i\dot{x}^j 
\end{equation}
respectively describe the free motion of particles and light rays. This is consistent as the Lagrangian itself is a constant of motion. The former coincide with the parameterization invariant extremals of the length functional $\int ds_R(x,dx)$, with 
\begin{equation}
\label{dsGR}
ds_{R}(x,dx)=\sqrt{g_{ij}(x)dx^idx^j} , 
\end{equation}
when these are parameterized by the same spacetime interval $ds_R(x,dx)$,
that is  physically interpreted as the proper time. 

It is worth emphasizing  that contrary to Riemannian geometry, where
$L_R(x,\dot{x})$ and $ds_R(x,dx)$ carry exactly the same information, in Lorentzian
geometry only  $L_R(x,\dot{x})$ is everywhere well defined and allows one to define
timelike and lightlike geodesics. The line element $ds_R(x,dx)$ makes sense only
within the light cones and does not allow a straightforward definition of
lightlike geodesics. For this reason in extending Finsler geometry to
indefinite signatures it seems to be preferable to focus on a Lagrangian that
is defined on the whole tangent bundle $TM$, rather than on the line
element that is  defined only over a subset of $TM$. Nonetheless, due to
the many complications that arise, both approaches have been considered and
widely discussed in the literature, producing a number of not always equivalent
definitions that can be roughly divided into two categories~\cite{LaemmerzahlandPerlick2018,Minguzzi2016}.

On the one hand, Asanov~\cite{Asanov1985} considered Finsler line elements
$ds_F(x,dx)$ that are positive-valued, non degenerate and positively homogeneous
of degree one in the second argument, defined only over a conical subset of the
tangent bundle $TM$, that is interpreted as the subbundle of future
directed timelike vectors.   The approach is appealing from the geometrical
point of view as it deals directly with the spacetime line element, but it is
not clear a priori how to describe the propagation of light rays as null
geodesics are not defined. The approach has been formalized and repeatedly
refined in particular by Javaloyes and S\'anchez~\cite{Javaloyes&Sanchez2014a,Javaloyes&Sanchez2014b,Javaloyes&Sanchez2020,BJS2020} by 
imposing conditions to retain the null directions. However, the increasingly complexity of the definition and the need to introduce by hand the desired causal structure by imposing conditions on the domain of $ds_F$ are not completely satisfactory from the physical point of view.

On the other hand, Beem~\cite{Beem1970} defined an indefinite Finsler
spacetime structure directly in terms of a real-valued Finsler Lagrangian
$L_F(x,\dot{x})$, defined and sufficiently differentiable on the whole spacetime
tangent bundle with the zero section removed $TM\backslash\{0\}$, that is homogeneous
of degree two in the fiber coordinates,
\begin{equation}
L_F(x,\lambda\dot{x})=\lambda^2L_F(x,\dot{x})\hskip10pt\text{for all $\lambda>0$},
\end{equation}
and such that 
\begin{equation}
\displaystyle{\bg_F}_{ij}(x,\dot{x})=\dpar_i\dpar_jL_F(x,\dot{x})
\end{equation}
is non degenerate with signature $+,-,-,-$. The spacetime structure is also called reversible if, in addition,
\begin{equation}
\label{reversibility}
L_F(x,-\dot{x})=L_F(x,\dot{x}).
\end{equation}
Due to Euler's homogeneous function theorem, the homogeneity of degree two in the second argument implies that the conserved Hamiltonian associated to $L_F$ coincides with the same Lagrangian, 
$H_F=L_F=\frac{1}{2}{\bg_F}_{ij}(x,\dot{x})\dot{x}^i\dot{x}^j$, so that at each spacetime point $x$, a direction
$\dot{x}$ can still  be classified as timelike, lightlike or spacelike
according to whether $H_F(x,\dot{x})$ is positive, zero or negative, exactly as in
the Lorentzian geometry. Reversibility guarantees that the spacetime
$M$ has the desired Lorentzian causal structure at any
point~\cite{Minguzzi2015}. As $H_F(x,\dot{x})$ is a constant of motion along the
affinely parameterized solutions of the associated Euler--Lagrange equations, we
can still interpret timelike and lightlike solutions as describing the free
motion of particles and light rays respectively. Again, within lightcones the
former coincide with the parameterization invariant extremals of the length
functional $\int ds_F(x,\dot{x})$, with 
\begin{equation}
ds_F(x,dx)=\sqrt{2L_F(x,dx)},
\end{equation}
when these are parameterized by the same spacetime interval $ds_F(x,dx)$, that is identified with the Finsler proper time.
Beem's definition is simple and quite satisfactory from the mathematical point
of view. However, it excludes several cases of physical interest where the
Lagrangian $L_F(x,\dot{x})$ fails to be well defined, such as for Randers
spaces~\cite{Randers1941}, or is not differentiable on a subset of measure
zero of $TM\backslash\{0\}$~\cite{LPH2012}. For this reason a number of refinements
have been put forward. In particular, L\"ammerzahl, Perlick and Hasse~\cite{LPH2012} required the Lagrangian to be defined and at least three times
continuously differentiable almost everywhere (that is, up to a set of measure
zero) on $TM\backslash\{0\}$   and the solutions of the associated Euler--Lagrange
equations to admit a continuous extension where the Finsler metric ${\bg_F}_{ij}(x,\dot{x})$
is not well defined; Pfeifer and Wohlfarth~\cite{PfeiferandWohlfarth2011,PfeiferandWohlfarth2012} also relaxed the regularity conditions,
introduced a `unit timelike condition' and extended the definition to
positively homogeneous Lagrangians of arbitrary degree; recently Hohmann,
Pfeifer and Voicu~\cite{HPV2019} further refined the definition in such a way
that the properties of the indefinite Finsler geometry are characterized by
four conic subbundles of $TM\backslash\{0\}$. 
The possibility of considering also non-homogeneous Lagrangians has also been
investigated by Miron and collaborators~\cite{Miron&Anastasiei1994}.

 Below we will consider yet another possible definition of spacetime structure. This is still in Beem's spirit, but it differs profoundly from it in the consideration that \textit{the demand for homogeneity is natural and necessary only for the line element $ds(x,dx)$ that describes the geometry of spacetime and not for the Lagrangian $L(x,\dot{x})$ that defines the casual structure of spacetime and its line element}. All that matters is that the timelike solutions of the Euler--Lagrange equations of $L(x,\dot{x})$ coincide with the parameterization invariant extremals of a length functional $\int ds(x,dx)$, when these are properly parameterized.
Consequently, we start from the general assumption that the free motion of elementary test particles in spacetime is described by an unspecified regular Lagrangian of which nothing is known except that in a certain limit it must correspond to (\ref{LGR}), a result that we will derive from the spacetime structure rather than impose it as a condition. 

We thus define a \textit{spacetime structure}  in terms of a real-valued Lagrangian function $L(x,\dot{x})$, defined, at least twice continuously differentiable in the base coordinates $x$ and analytic in the fiber coordinates $\dot{x}$ on the whole tangent bundle $TM$ (the zero section is not removed) which has the following properties:
\begin{itemize}
  \item[(i)] $L(x,0)=0$;
  \item[(ii)]  $\bg_{ij}(x,\dot{x})\equiv
\dpar_i\dpar_jL (x,\dot{x})$ is non degenerate with signature $+,-,-,-$.
\end{itemize}
As for Beem's, the spacetime structure is also called reversible if, in
addition,  (\ref{reversibility}) is satisfied. The condition (i) is
physically interpreted as the absence of the potential energy of external force
fields, thus guaranteeing that the energy of test particle is purely kinetic.
The condition (ii), as customary, guarantees the existence and
uniqueness of the solutions of the associated Euler--Lagrange equations, that
$\bg_{ij}(x,\dot{x})$ has constant Lorentzian signature on $TM$ and that the
spacetime signature is the desired one. Clearly, the definition
straightforwardly extends to any spacetime dimension and signature. It is also
worth emphasizing that this definition does not reduce to any of those
discussed above in the case of homogeneous Lagrangians, because the zero
section is not removed from the tangent bundle. As an example, Lagrangians of
the type $L=G_{i_1...i_n}\dot{x}^{i_1}...\dot{x}^{i_n}$, with $n>2$~\cite{Punzietal.2009},  do not fit in the definition, because $\dpar_i\dpar_jL$ is degenerate on the zero section.
 
As the Lagrangian does not explicitly depend on the evolution parameter, the corresponding Euler--Lagrange equations admit, due to the first Noether theorem, a first integral of the energy type
\begin{equation}
\label{energy}
H=\dot{x}^i\dpar_iL-L.
\end{equation}
Accordingly, at each spacetime point $x$ a direction $\dot{x}$ is classified as timelike, lightlike or spacelike according to whether $H(x,\dot{x})$ is positive, zero or negative, and the timelike and lightlike solutions of the Euler--Lagrange equations of $L(x,\dot{x})$ are respectively interpreted as the timelike and lightlike geodesics  describing the free motion of particles and light rays. Quite remarkably, for each positive value of the energy-type first integral $H$, the former coincide with the  extremals of the length functional $\int ds_{J}(x,dx)$,  where $ds_{J}(x,dx)$ is the Jacobi metric for the Lagrangian $L(x,\dot{x})$.

\section{The Jacobi metric}
\label{The Jacobi metric} 
It has long been known~\cite{Jacobi1842-43,Goldstein1970,AKN2006} that for
each fixed energy value $E$, the paths in configuration space of a
`natural' Lagrangian system
\begin{equation}
L=\frac{1}{2}g_{ij}(x)\dot{x}^i\dot{x}^j-V(x),
\end{equation} 
are the geodesics of the energy-dependent Jacobi metric of Riemannian type
\begin{equation}
\label{dsJ}
ds_J(x,\dot{x},E)=\sqrt{2[E-V(x)]g_{ij}(x)dx^idx^j}.
\end{equation}
This result, together with its extension to relativistic Lagrangians, has a
number of  applications  in mathematical physics and general
relativity~\cite{Levi-Civita1917,Weyl1917,Perlick1991,SHS1996,Gibbons2016,CGGMW2019} and for $V(x)=0$ justifies the equality of the positive
energy world lines of the Lagrangian
(\ref{LGR}) with the geodesics of the spacetime line element (\ref{dsGR}). Of
particular importance to us, Ba\.{z}a\'nski and Jaranowski~\cite{Bazanski&Jaranowski1994} have generalized this result to arbitrary Lagrangian systems by showing that,  for each fixed energy value $E$, the paths in
configuration space of any non-degenerate Lagrangian $L(x,\dot{x})$, are the
geodesics of an energy-dependent Finsler line element $ds_{J}(x,dx,E)$ (see
also~\cite{Bazanski2003,Mestdag2016,Maraner2019}). Therefore, to every
regular Lagrangian corresponds a one-parameter family of Finsler geometries.
Ba\.{z}a\'nski and Jaranowski also posed and solved the inverse problem by showing that given any Finsler line element $ds_F(x,dx)$ and an energy function $H(x,\dot{x})$, it is always possible to find a non-degenerate Lagrangian $L(x,\dot{x})$ whose associated Jacobi line element is, for a specific energy value, the given line element. Therefore, every Finsler geometry can be described, in countless different ways, depending on the choice of $H(x,\dot{x})$ and $E$, in terms of a regular Lagrangian that is not necessarily homogeneous in the $\dot{x}$.

Hereafter, we will use the hypothesis that the Lagrangian defining the spacetime structure is analytic in the fiber coordinates to series expand it about $\dot{x}=0$. This will allow us to set up a spacetime formalism that directly connects and extends that of general relativity and classical field theory. In fact, by taking into account $(\text{i})$, we have that
\begin{equation}
\label{L}
L=l_i(x)\dot{x}^i+\frac{1}{2}l_{ij}(x)\dot{x}^i\dot{x}^j+\cdots+\frac{1}{n!}l_{i_1...i_n}(x)\dot{x}^{i_1}...\ \dot{x}^{i_n}+\cdots
\end{equation}
where the
\begin{equation}
l_{i_1...i_n}(x)=\dpar_{i_1}...\dpar_{i_n}L(x,0)
\end{equation}
transform as symmetric tensors for all $n>0$ and fully characterize the geometry of spacetime. We observe that the set of fields $l_{i_1...i_n}$ is completely equivalent to the assignment of the Lagrangian, so that we will indifferently refer to them  as the spacetime Lagrangian.
If the Lagrangian is required to be reversible, all odd rank tensors, $l_i, l_{ijk}, \ldots $, are identically vanishing. In the quadratic approximation
(\ref{L}) reduces to (\ref{LGR}) with $l_{ij}$ identified with the spacetime metric $g_{ij}$, and the possible addition of a linear term $l_i$, that will be identified with the electromagnetic potential. For this reason, it is important to emphasize that the required non degeneracy of $\bg(x,\dot{x})=l_{ij}(x)+ l_{ijk}(x)\dot{x}^k+\frac{1}{2} l_{ijkl}(x)\dot{x}^k\dot{x}^l+\cdots$ on the whole tangent bundle, zero section included, implies the non degeneracy
of $l_{ij}(x)$. 

The energy type first integral  (\ref{energy}) defining the casual structure of spacetime is evaluated as 
\begin{equation}
\label{H}
H=\frac{1}{2}l_{ij}(x)\dot{x}^i\dot{x}^j+\cdots+\frac{n-1}{n!}l_{i_1...i_n}(x)\dot{x}^{i_1}...\ \dot{x}^{i_n}+\cdots,
\end{equation}
and attains a constant value on the trajectories of the system that, when positive, will be identified with half of the square of the invariant mass of the test particle.
We observe that, as far as the fields $l_{i_1...i_n}(x)$ with $n>2$ are somehow small compared to $l_{ij}(x)$, the casual structure of  spacetime is just a small deformation of that  determined by the quadratic form $l_{ij}(x)\dot{x}^i\dot{x}^j$.

For trajectories corresponding to positive values of $H$, thus for timelike trajectories,  
the Jacobi metric for the general Lagrangian (\ref{L}) has been evaluated
in~\cite{Maraner2019} as
\begin{equation}
\label{dsJ}
\begin{split}
ds_{J}=&l_i(x)dx^i+\mu\sqrt{l_{ij}(x)dx^idx^j}+\mu^2\frac{1}{6}\frac{l_{ijk}(x)dx^idx^jdx^k}{l_{ij}(x)dx^idx^j}\\
&+\mu^3\left[\frac{1}{24}\frac{l_{ijkl}(x)dx^i dx^j dx^k dx^l}{\left(l_{ij}(x)dx^i dx^j\right)^{\frac{3}{2}}}-\frac{1}{18}\frac{\left(l_{ijk}(x)dx^i dx^j dx^k\right)^2}{\left(l_{ij}(x)dx^i dx^j\right)^{\frac{5}{2}}}\right]+\cdots,
\end{split}
\end{equation}
where we have set $2H=\mu^2$. 
Beyond the formal manipulations that lead to this expression,
the validity of (\ref{dsJ}) is proven \textit{a posteriori} by the fact that
for each value of $\mu$ the geodesic equations for the line element
$ds_{J}$, when appropriately parameterized, are identical to the
Euler--Lagrange equations for the Lagrangian $L$ (see below).
Correspondingly, the causal structures of spacetime induced by $L$ and
$ds_{J}$ are identical. A particular case of (\ref{dsJ}) has recently been
considered in~\cite{LoboandPfeifer2020}.

The Jacobi line elements $ds_J$ are defined only on the subset 
of the tangent bundle $\{(x,\dot{x})\in TM\ \vert\ H>0\}$ that is interpreted as the subbundle of future directed timelike vectors. For small values of  $\mu$, this set is indistinguishable from $\{(x,\dot{x})\in TM\ \vert\ l_{ij}(x)\dot{x}^i\dot{x}^j>0\}$, so that in this regime the causal structure of spacetime results indistinguishable from that of general relativity.

In the reversible case the Jacobi line element appears as a generalization of
the Lorentzian line element (\ref{dsGR})  and in the non reversible case as a
generalization of the Randers line element~\cite{Randers1941}. 
In the context of Lagrangian mechanics this structure is universal. The fields content and the explicit form of the line element emerge spontaneously from the assumption that the motion of test particles is described by an unspecified free regular Lagrangian, without the need of being postulated. In the limit of small values of $\mu$, the trajectories of motion corresponding to positive values of the energy type first integral 
of \textit{any} regular Lagrangian system are therefore approximated by the geodesics of an energy dependent Randers metric or, equivalently, by the paths in configuration space of a representative point moving under the action of force fields of the electromagnetic and gravitational type.

The differential of the Lagrangian evolution parameter $t$ can also be
expressed in terms of the tensors $l_{i_1...i_n}(x)$ and the differentials
$dx^i$~\cite{Maraner2019} as
\begin{equation}
\label{Lt}
\begin{split}
dt=&\frac{1}{\mu}\sqrt{l_{ij}(x)dx^idx^j}+\frac{1}{3}\frac{l_{ijk}(x)dx^idx^jdx^k}{l_{ij}(x)dx^idx^j}
\\
&+\mu\left[\frac{1}{8}\frac{l_{ijkl}(x)dx^i dx^j dx^k dx^l}{\left(l_{ij}(x)dx^i dx^j\right)^{\frac{3}{2}}}-\frac{1}{6}\frac{\left(l_{ijk}(x)dx^i dx^j dx^k\right)^2}{\left(l_{ij}(x)dx^i dx^j\right)^{\frac{5}{2}}}\right]+\cdots
\end{split}
\end{equation}
and is related to the Jacobi line element (\ref{dsJ}) by the identity $dt=\mu\frac{d}{d\mu}ds_{J}$.
Up to a constant rescaling by a factor $1/\mu$, it coincides in first approximation with the differential of the proper time of General Relativity. Therefore, the small $\mu$ regime of the theory  indicates that  we must identify the differential of the proper time along timelike world lines with 
\begin{equation}
\label{pt}
d\tau=\mu dt=\mu^2\frac{d}{d\mu}ds_{J}.
\end{equation}
We will call $\tau$ the proper time of the particle with energy type first
integral $H=\mu^2/2$. It is worth emphasizing that this definition of the
proper time is equivalent to the one from the clock postulate~\cite{Pfeifer2019} only in the small $\mu$ regime.

\section{Spacetime geometry}
\label{Spacetime geometry}
As in Riemannian geometry, the investigation of the metric relations of spacetime is approached through the study of the geodesics, the lines extremizing the length functional 
\begin{equation}
\int ds_{J}(x,dx)=\int ds_{J}\left(x,{\textstyle\frac{dx}{d\theta}}\right)d\theta,
\end{equation}
where, as usual, we have taken advantage of the homogeneity of degree one in the second argument of the line element to introduce an arbitrary parameterization. 

\subsection{Geodesic equations}

The direct variation of the length functional produces the geodesic equations in the form
\begin{equation}\label{ge}
\begin{split}
&\Gamma_{hi}\acute{x}^i
\\ &\hskip10pt 
+\frac{1}{\acute{t}}\left(l_{hi} \text{\textit{\H{x}}}^i+\Gamma_{hij}\acute{x}^i\acute{x}^j\right)-\frac{\text{\textit{\H{t}}}}{\acute{t}^{\,2}}l_{hi}\acute{x}^i
\\ &\hskip20pt  
+\frac{1}{\acute{t}^{\,2}}\left(l_{hij} \text{\textit{\H{x}}}^i\acute{x}^j+\Gamma_{hijk}\acute{x}^i\acute{x}^j\acute{x}^k\right)-\frac{\text{\textit{\H{t}}}}{\acute{t}^{\,3}}l_{hij}\acute{x}^i\acute{x}^j
\\ &\hskip30pt 
+\frac{1}{\acute{t}^{\,3}}\left(\frac{1}{2}l_{hijk} \text{\textit{\H{x}}}^i\acute{x}^j\acute{x}^k+\Gamma_{hijkl}\acute{x}^i\acute{x}^j\acute{x}^k\acute{x}^l\right) -\frac{1}{2}\frac{\text{\textit{\H{t}}}}{{\acute{t}}^{\, 4}}l_{hijk}\acute{x}^i\acute{x}^j\acute{x}^k+...=0,
\end{split}
\end{equation}
where the acute accent indicates the differentiation with respect to the parameter $\theta$, $\acute{x}=\frac{dx}{d\theta}$, $t$ is the Lagrangian evolution parameter as expressed in (\ref{Lt}) and the quantities $\Gamma_{hi_1i_2...i_n}(x)$ are defined as
\begin{equation}
\label{Gammas}
\begin{split}
\Gamma_{hi} =&\ \partial_i l_h-\partial_h l_i, \\
\Gamma_{hij}  =&\ \frac{1}{2}\left(\partial_i l_{hj}
+\partial_j l_{ih}-\partial_h l_{ij}\right), \\
\Gamma_{hijk}  =&\ \frac{1}{3!}\left(\partial_i l_{hjk}
+\partial_j l_{ihk}+\partial_k l_{ijh}
-\partial_h l_{ijk}\right),  
\\
\Gamma_{hijkl}  =&\ \frac{1}{4!}\left(\partial_i l_{hjkl}
+\partial_j l_{ihkl}+\partial_k l_{ijhl}+\partial_l l_{ijkh}
-\partial_h l_{ijkl}\right), \\
&  ...
\end{split}
\end{equation}
The quantities $\Gamma_{hi_1...i_n}(x)$ do not transform as spacetime tensors and appear as a generalization of both, the force  field for the vector potential $l_i$ and the Christoffel symbols of the first kind for the metric tensor $l_{ij}$. 

It is clear from (\ref{ge}), that the geodesic equations take their simpler form when the parameter $\theta$ is chosen directly proportional to the Lagrangian evolution parameter $t$. 
This choice, that explicitly breaks the invariance of the equations under a constant rescaling of the parameter,  is not customary in Finsler geometry where the arc length  is used instead.  
In physics, however, the evolution parameter has a precise observable meaning 
and it has to reduce to the proper time in the general relativistic limit, a property not enjoyed by the arc length $ds_{J}$ in the non reversible case. We thus proceed by choosing the evolution parameter as the proper time (\ref{pt}). For each value of $\mu$
the geodesic equations  take then the simple form 
\begin{equation}
\label{ELe}
\bg_{hi}{x''}^i+\frac{1}{\mu}\Gamma_{hi}{x'}^i+\Gamma_{hij}{x'}^i{x'}^j+\mu\Gamma_{hijk}{x'}^i{x'}^j{x'}^k+\mu^2\Gamma_{hijkl}{x'}^i{x'}^j{x'}^k{x'}^l+...=0,
\end{equation}
where the prime indicates the differentiation with respect to $\tau$, ${x'}=\frac{dx}{d\tau}$, and 
\begin{equation}
\label{bg}
\bg_{hi}=l_{hi}+\mu l_{hij}{x'}^j+\frac{1}{2}\mu^2 l_{hijk}{x'}^j{x'}^k+\cdots
\end{equation}
is defined in $(\text{ii})$.
Equations (\ref{ELe}) appear as a straightforward generalization of the covariant form of the geodesic equations of Riemannian geometry. By introducing the tensor fields
\begin{equation}
\label{F}
F_{hi_1...i_n}=\frac{1}{n!}\left(\nabla_{h}l_{i_1...i_{n}}-\nabla_{i_1}l_{hi_2...i_{n}}-\cdots-\nabla_{i_n}l_{i_1...i_{n-1}h}\right),
\end{equation}
where $\nabla_k$ denotes the covariant derivative induced by the metric $l_{ij}$, the  equations can be given the  explicit covariant form 
\begin{equation}
\label{cELe}
\bg_{hk}\left({x''}^k+{\Gamma^k}_{ij}{x'}^i{x'}^j\right)-\frac{1}{\mu}F_{hi}{x'}^i-\mu F_{hijk}{x'}^i{x'}^j{x'}^k-\mu^2F_{hijkl}{x'}^i{x'}^j{x'}^k{x'}^l+\cdots=0,
\end{equation}
with ${\Gamma^k}_{ij}=l^{kh}\Gamma_{hij}$ denoting the Christoffel symbols of the second kind for the metric $l_{ij}$. 
Up to a constant rescaling of the evolution parameter, these equations, obtained under the assumption that $H>0$, coincide with the Euler--Lagrange equations for the spacetime Lagrangian (\ref{L}). Correspondingly, in the reversible case, where the term $\frac{1}{\mu}F_{hi}{x'}^i$ as well as all terms $\mu^{2n-1}F_{hi_1...i_{2n+1}}\dot{x}^{i_1}...\dot{x}^{i_{2n+1}}$ are absent, these equations can be used to describe the motion of particles with  $H=0$, as done in general relativity. The limit $\mu\to0$ does not cause any problem in this case.

The geodesic equations for the Finsler geometries that can be described by the line element (\ref{dsJ}) can thus be put in correspondence with the equations of motion of a charged particle with mass-to-charge ratio proportional to $\mu$,  moving in the Riemannian background metric $ l_{ij}$ under the action of a potentially infinite number of tensorial force fields $F_{hi_1...i_n}$, that straightforwardly generalize the electromagnetic and gravitational interactions.
 For small values of $\mu$ these interactions act hierarchically, with the  electromagnetic one more intense and those of a greater rank that become gradually weaker. 
 
To check whether in spacetime physics the lower-rank potentials $l_i$ and $l_{ij}$ can tentatively be identified with real electromagnetic and gravitational potentials it is necessary to introduce dimensional quantities. By taking into account the relative strength of electromagnetic and gravitational forces we thus set  
\begin{equation}
l_i\equiv\frac{\sqrt{4\pi\varepsilon_0G}}{c}A_i
\hskip10pt\text{and}\hskip10pt
 l_{ij}\equiv g_{ij}
\end{equation}
with  $\varepsilon_0$ the permittivity of free space, $G$ the gravitational constant, $c$ the speed of light, $A_i$ the electromagnetic vector potential and $g_{ij}$ the spacetime metric. Correspondingly, by comparing (\ref{cELe}) with the equations of motion of an elementary test particle of invariant mass $m_0$ and charge the elementary charge $e$, the adimensional parameter $\mu$ results in 
\begin{equation}
\mu=\pm\frac{m_0}{m_S}=\pm\frac{1}{\sqrt{\alpha}}\frac{m_0}{m_P}
\end{equation}
with $m_S=\sqrt{\frac{e^2}{4\pi\varepsilon_0G}}$ the Stoney mass, $\alpha$ the fine structure constant and $m_P$ the Planck mass.
For elementary test particles with invariant mass of the order of electron and proton we respectively obtain 
\begin{equation}
|\mu_{electron}|\approx10^{-21}
\hskip10pt\text{and}\hskip10pt
|\mu_{proton}|\approx10^{-18},
\end{equation}
amply justifying the fact that the effects of the interactions of rank three and higher have never been observed in experiments. 
Neutral particles like neutrinos and neutrons are accommodated in the theory by postulating that they are described by reflexive Lagrangians, in such a way that all odd rank interactions cancel out. Correspondingly, the trajectories of photons are described  as in general relativity as the $\mu\to0$ limit of equations~(\ref{cELe}) with all odd rank interaction set equal to zero, with the evolution parameter no longer corresponding to the proper time, that vanishes, but directly determined by the equations. 
Therefore, with all the limitations of a classical theory, the idea that the electromagnetic and gravitational interactions that we experience emerge independently of the specific dynamics chosen for  `classical elementary particles' seems to be at least worth of deeper investigation.

Given the postulated non-degeneracy of $\bg_{ij}$, the geodesic equations can also be resolved with respect to the second derivatives by contraction with the inverse $\bg^{ij}$ of $\bg_{ij}$, $\bg^{ih}\bg_{hj}=\delta^i_j$. The existence and uniqueness for their solutions follow then from the standard theory of ordinary differential equations.  After renaming indices, we obtain the geodesics equations in the form  
\begin{equation}
\label{geq}
\begin{split}
{x''}^h&-\frac{1}{\mu}{F^h}_{i}{x'}^i
+\Bigl({\Gamma^h}_{ij}+{l^h}_{li}{F^l}_{j}\Bigr){x'}^i{x'}^j\\
&+ \mu\Bigl(-{F^h}_{ijk}+{l^h}_{li}{l^{lm}}_jF_{km}+\frac{1}{2}{l^h}_{lij}{F^l}_{k}\Bigr){x'}^i{x'}^j{x'}^k+\cdots=0
\end{split}
\end{equation}
where indices have been raised by means of the inverse of $g_{ij}$. When  reparameterized by the Lagrangian evolution parameter $t=\frac{1}{\mu}\tau$, these equations all take the form of the Euler--Lagrange equations for the Lagrangian (\ref{L})
\begin{equation}
\label{geqL}
\begin{split}
\ddot{x}^h+\Big[-&{F^h}_{i}
+\Big({\Gamma^h}_{ij}+{l^h}_{li}{F^l}_{j}\Big)\dot{x}^j\\
&+\Big(-{F^h}_{ijk}+{l^h}_{li}{l^{lm}}_jF_{km}+\frac{1}{2}{l^h}_{lij}{F^l}_{k}\Big)\dot{x}^j\dot{x}^k+...\Big]\dot{x}^i=0,
\end{split}
\end{equation}
with the dot now indicating differentiation with respect to $t$, $\dot{x}=\frac{dx}{dt}$.  
While the parameterization in terms of the proper time $\tau$ is crucial in describing particles dynamics, in the investigation of spacetime geometry we find it more convenient to start from the Lagrangian parameterization (\ref{geqL}) where all geodesic equations take the same form. It should be kept in mind, however, that 
although the parameter $\mu$ no longer explicitly appear in the equations,  the solutions that connect two given spacetime points are different for different values of the energy-type first integral $H=\mu^2/2$.
Solutions corresponding to different values of $\mu$ describe different geometries.

\subsection{Connection and curvature}

For each value of the parameter $\mu$ the geodesic equations~(\ref{geqL}) can be read as
the auto parallel equations~\cite{BCZ2012,Minguzzi2014,Miron&Anastasiei1994}
\begin{equation}
\label{apEq}
\ddot{x}^h+{\bgamma^h}_i\dot{x}^i=0
\end{equation}
for the $\mu$ dependent non-linear connection 
\begin{equation}
\begin{split}
{\bgamma^h}_i=-&{F^h}_{i}
+\Big[{\Gamma^h}_{ij}+{l^h}_{l(i}{F^l}_{j)}\Big]\dot{x}^j\\
&+\Big[-{F^h}_{ijk}+{l^h}_{l(i}{l^{lm}}_jF_{k)m}+\frac{1}{2}{l^h}_{l(ij}{F^l}_{k)}\Big]\dot{x}^j\dot{x}^k+...+{\bchi^h}_i,
\end{split}
\end{equation}
where  round brackets indicate symmetrization and ${\bchi^h}_i$ is any quantity that transforms like a spacetime tensor and vanishes when its second index is contracted with $\dot{x}^i$, ${\bchi^h}_i\dot{x}^i=0$. 
Indeed, it should be emphasized that there is a margin of freedom in extracting a connection from the geodesic equations.  This freedom is already present in the Riemannian geometry, where it allows for the introduction of the torsion.
It is however more pronounced in the context of Finsler and Lagrange geometries, where a non vanishing ${\bchi^h}_i$ does not require the introduction of new fields and can be built in terms of quantities that already appear in the line element.  For example, a term like $f(F^{kl}F_{kl})\dot{x}^hF_{ij}\dot{x}^j$, with $f$ an arbitrary function of $F^{kl}F_{kl}$, can be added to the connection without modifying the auto parallel equations. 
In Riemannian geometry, where  the geodesic equations read 
\begin{equation}
\ddot{x}^h+{\Gamma^h}_{ij}\dot{x}^i\dot{x}^j=0,
\end{equation}
the (Levi-Civita) connection is chosen as 
\begin{equation}
\label{L-C}
{\bgamma^h}_i=\frac{1}{2}\dpar_i\left({\Gamma^h}_{jk}\dot{x}^j\dot{x}^k\right)= {\Gamma^h}_{ij}\dot{x}^j,
\end{equation}
thus with ${\bchi^h}_i=0$ and is the only connection that preserves the metric tensor $g_{ij}$ and it is torsion free. 
In Finsler geometry, where the geodesic equations read 
\begin{equation}
\ddot{x}^h+\bg_F^{hk}\left(\partial_i\dpar_kL_F\dot{x}^i-\partial_kL_F\right)=0,
\end{equation}
the (Cartan) non linear connection is chosen as 
\begin{equation}
\label{Cartan}
{\bgamma^h}_i\equiv{N^h}_i=\frac{1}{2}\dpar_i\left[\bg_F^{hk}\left(\partial_i\dpar_kL_F\dot{x}^i-\partial_kL_F\right)\right],
\end{equation}
thus with a non vanishing ${\bchi^h}_i$ and corresponds again to the only
connection that preserves the metric tensor $\bg_{Fij}$ and it is torsion
free~\cite{BCZ2012}. However, it can also be characterized  as the only non
linear connection constructed solely in terms of the Finsler Lagrangian
$L_F$ that leaves  the associated conserved Hamiltonian (that due to
Euler's Homogeneous Function Theorem also corresponds to the Lagrangian itself,
$H_F=\dot{x}^i\dpar_iL_F-L_F=L_F$) covariantly constant,
\begin{equation}
\delta_iH_F\equiv (\partial_i-{\bgamma^h}_i\dpar_h)H_F=\delta_iL_F=0.
\end{equation}  
In Miron's Lagrange geometry~\cite{Miron&Anastasiei1994} the connection is
chosen identical in form to the Cartan connection (\ref{Cartan}), with the
drawback that its auto parallel equations no longer correspond the original
Euler--Lagrange equations. For this reason we find it more appropriate to
choose the connection as the only non linear connection constructed solely in
terms of the Lagrangian $L$ that leaves the corresponding conserved
Hamiltonian $H$ covariantly constant 
\begin{equation}
\label{dH=0}
\delta_i H\equiv (\partial_i-{\bgamma^h}_i\dpar_h)H=0
\end{equation}
and whose auto parallel equations exactly correspond to the Euler--Lagrange equations~(\ref{geqL}).
This general condition is completely geometrical in character and seems to us the natural generalization of the choices operated in Riemannian and Finsler geometry. 
In particular, in the Riemannian geometry, where ${\bgamma^h}_i$ is given by \ref{L-C} and $H=\frac{1}{2}g_{ij}\dot{x}^i\dot{x}^j$, 
the condition reduces to $\left(\partial_k g_{ij}-{\Gamma^h}_{ki}g_{hj}-{\Gamma^h}_{kj}g_{ih}\right)\dot{x}^i\dot{x}^j=0$ that, together with the requirement that the connection depend only on $g_{ij}$, directly leads to (\ref{L-C}). The explicit form of the connection for more general spacetime geometries will be worked out here only in the small $\mu$ regime of the theory.

Regardless of the choice of the connection, the commutator of the horizontal
derivatives  $\delta_i =\partial_i-{\bgamma^h}_i\dpar_h$, $[\delta_i,\delta_j]={\br^h}_{ij}\dpar_h$, defines the curvature tensor~\cite{BCZ2012,Miron&Anastasiei1994,Minguzzi2014}
\begin{equation}
\label{brhij}
{\br^h}_{ij}=\partial_i{\bgamma^h}_j-\partial_j{\bgamma^h}_i-
{\bgamma^k}_i\dpar_k{\bgamma^h}_j+
{\bgamma^k}_j\dpar_k{\bgamma^h}_i.
\end{equation}  
In Riemannian geometry this reduces to ${\br^h}_{ij}={R^h}_{kij}\dot{x}^k$ with ${R^h}_{kij}=\partial_i{\Gamma^h}_{jk}-\partial_j{\Gamma^h}_{ik}-{\Gamma^l}_{ik}{\Gamma^h}_{jl}+{\Gamma^l}_{jk}{\Gamma^h}_{il}$ the Riemann curvature tensor.
By contraction of a covariant and a contravariant index of ${\br^h}_{ij}$ it is possible to construct the contracted curvature tensor 
\begin{equation}
{\br}_{i}={\br^h}_{hi},
\end{equation}
that in the Riemannian geometry corresponds to  ${\br}_{i}=R_{ij}\dot{x}^j$ with $R_{ij}=
{R^k}_{ikj}$ the Ricci tensor.
Finally, by contraction with $\dot{x}^i$ it is possible to construct the curvature scalar
\begin{equation}
\label{br}
\br={\br}_{i}\dot{x}^i,
\end{equation}
that is relevant for the tidal acceleration of Riemannian and Finsler
geodesics~\cite{Rutz1993} but does not reduce to the scalar curvature
$R=g^{ij}R_{ij}$ in the Riemannian geometry. Pfeifer and Wohlfarth~\cite{PfeiferandWohlfarth2011,PfeiferandWohlfarth2012} suggested using this scalar as
the Lagrangian density in Finsler gravity.

\subsection{Geometry of the small $\mu$ regime}
We are now ready to investigate the geometry of the configuration space in the regime of small values of the energy type first integral $H=\mu^2 /2$. Since in spacetime physics $|\mu|\lesssim10^{-18}$, this regime is largely justified and it is actually hard to imagine how to test the non-perturbative regime of the theory. For such values of $\mu$ the spacetime line element (\ref{dsJ}) is indistinguishable from the Randers line element
\begin{equation}
\label{dsJcft}
ds_{J}\approx l_i(x)dx^i+\mu\sqrt{g_{ij}(x)dx^idx^j}
\end{equation}
and, correspondingly, the series expansion (\ref{L}) of any regular free Lagrangian can be truncated to the second order
\begin{equation}
\label{L2}
L\approx l_i\dot{x}^i+\frac{1}{2}g_{ij}\dot{x}^i\dot{x}^j.
\end{equation}
We remark that this is not the choice of a particular Lagrangian.
 On the contrary, it is a universal feature of the small $\mu$ regime of the postulated spacetime structure. In this regime the tensor $\bg_{ij}$ reduces then to $g_{ij}$ and the Hamiltonian to $H=\frac{1}{2}g_{ij}\dot{x}^i\dot{x}^j$. Correspondingly,  the proper time (\ref{pt}) correctly reduces to the general relativistic proper time, $d\tau=\sqrt{g_{ij}dx^idx^j}$.
The geodesic equations for every $ds_{J}$ are then the equations of motion for  charged particles with mass-to-charge ratio proportional to $\mu$, moving in the background electromagnetic field $F_{ij}$ and metric $g_{ij}$. 
As before, to discuss the geometry of spacetime we reparametrize them by the Lagrangian evolution parameter $t=\frac{1}{\mu} \tau$,  so that they all have the same form
\begin{equation}
\ddot{x}^h-{F^h}_i\dot{x}^i+{\Gamma^h}_{ij}\dot{x}^i\dot{x}^j=0,
\end{equation}
corresponding to the Euler--Lagrange equations for the quadratic Lagrangian (\ref{L2}). 
The non linear spacetime connection is thus given by 
${\bgamma^h}_i=-{F^h}_i+{\Gamma^h}_{ij}\dot{x}^j+{\bchi^h}_i$, with ${\bchi^h}_i$ determined by the condition~(\ref{dH=0}) exclusively in terms of $l_i$ and $g_{ij}$. The corresponding equation has the unique solution ${\bchi^h}_i=-\frac{1}{2H} \dot{x}^hF_{ij}\dot{x}^j$. We thus obtain the genuinely non linear $\mu$ dependent spacetime connection as
\begin{equation}
\label{nlc} 
{\bgamma^h}_i=-{F^h}_i+  
{\Gamma^h}_{ij}\dot{x}^j-\frac{1}{2H} \dot{x}^hF_{ij}\dot{x}^j.
\end{equation}
It is worth emphasizing that this non linear connection is univocally determined by the Lagrangian $L$ and the geometrical condition~(\ref{dH=0}) and, to the best of our knowledge, has never been considered before in the literature. The corresponding  curvature tensor results in
\begin{equation}\label{brhij2}
\begin{split}
{\br^h}_{ij}=&-\nabla_i{F^h}_j+\nabla_j{F^h}_i+\left[{R^h}_{kij}-\frac{1}{2H}
\left({F^h}_iF_{jk}-{F^h}_jF_{ik}\right) \right]\dot{x}^k\\
&\hskip150pt-\frac{1}{2H}\dot{x}^h\left(\nabla_iF_{jk}-\nabla_jF_{ik}\right)\dot{x}^k
\end{split}
\end{equation}
and the associated curvature scalar (\ref{br})  reads
\begin{equation}
\label{bR2}
\br=-\nabla_j{F^j}_i\dot{x}^i+\left( R_{ij}+\frac{1}{2H}{F^k}_iF_{kj}\right)\dot{x}^i\dot{x}^j.
\end{equation}
The curvature is singular on the light cone $H=0$ and is dominated by the quadratic term in  $F_{ij}$ for small values of $H$.

\section{Spacetime dynamics}
\label{Spacetime dynamics}
The geometry of spacetime is determined by the scalar field  $L(x,\dot{x})$, identified with  its expansion coefficients $l_{i_1...i_n}(x)$,  on the tangent bundle $TM$. 
The dynamics of spacetime must therefore be described by an action principle on
the  spacetime tangent bundle for this scalar field~\cite{PfeiferandWohlfarth2012,NiandShen2019}. To specify it, it is necessary to provide a Lagrangian
density for $L(x,\dot{x})$ and an integration  measure for $TM$. 
Following Pfeifer and Wohlfarth~\cite{PfeiferandWohlfarth2012}, the former
is chosen as the curvature scalar (\ref{br}) and the latter as the volume
element associated to the Sasaki type tangent bundle metric
\begin{equation}
ds_{TM}^2=\bg_{ij}dx^idx^j+\bmu^2\bg_{ij}\left(d\dot{x}^i+{\bgamma^i}_kdx^k\right)
\left(d\dot{x}^j+{\bgamma^j}_kdx^l\right),
\end{equation}
corresponding to $\sqrt{\bmu^2\bg^2} d^4xd^4\dot{x}=|\bmu |\,|\bg|  
d^4xd^4\dot{x}$, with $\bg=\det\bg_{ij}$. The function  $\bmu$ is chosen by observing that the integration over the whole tangent bundle implies an averaging over states with different  $\mu$. Since the logarithm of the probability distribution of a dynamical system is an additive constant of motion and the only additive constant of motion of the system is the Hamiltonian, it is reasonable to expect  this averaging to be weighted by a Gibbs factor.  We therefore set  
\begin{equation}
\bmu(x,\dot{x})=\frac{1}{ \mathcal{N}(\kappa)}e^{-\kappa H(x,\dot{x})},
\end{equation}
 with $H(x,\dot{x})$ the Hamiltonian (\ref{H}), $\kappa$ a constant and $\mathcal{N}(\kappa)$ an appropriate, possible infinite, $\kappa$-dependent normalization constant.  

Coupling to matter can be considered consistently by means of a Lagrangian density $\bL_m(x,\dot{x})$ on $TM$ which describes the energy-moment distribution of matter in a given geometric configuration and that we assume to be sufficiently regular to be series expanded about $\dot{x}=0$,
\begin{equation}
\label{bLm}
\bL_m=\mathcal{L}(x)+\mathcal{L}_i(x)\dot{x}^i+\frac{1}{2}\mathcal{L}_{ij}(x)\dot{x}^i\dot{x}^j+\cdots+\frac{1}{n!}\mathcal{L}_{i_1...i_n}(x)\dot{x}^{i_1}...\ \dot{x}^{i_n}+\cdots
\end{equation}
As for the geometry, the coefficients of this expansion will play the role of the spacetime tensors characterizing the distribution of matter.

The full action for the dynamics of spacetime is thus chosen as
\begin{equation}
\label{S}
S[L]=-\frac{1}{\mathcal{N}(\kappa)}\iint_{TM} \left(\br+\bL_m\right) e^{-\kappa H}| \bg| 
d^4xd^4\dot{x},
\end{equation}
where the integration is taken over the whole tangent bundle $TM$. 
We remark that, if the Lagrangian $L(x,\dot{x})$ satisfies the condition (i) and is analytic in the fiber coordinates $\dot{x}$, the fields $l_{i_1...i_n}(x)$ are completely determined by $L$ and, vice versa, they completely determine $L$. Correspondingly, 
\begin{equation}
L(x,\dot{x})\hskip5pt\text{and}\hskip5pt \Bigl\{l_i(x)\sim A_i(x),\hskip5pt l_{ij}(x)=g_{ij}(x),\hskip5pt l_{ijk}(x),\hskip5pt l_{ijkl}(x),\hskip5pt ...\Bigr\}
\end{equation}
are just two different, but equivalent, ways of describing the same tangent bundle scalar field. By the chain rule, the variation of any sufficiently regular functional of $L$ can thus be obtained as 
\begin{equation}
\frac{\delta S[L]}{\delta L}=
\frac{\delta S[L]}{\delta l_i}\delta l_i+
\frac{\delta S[L]}{\delta l_{ij}}\delta l_{ij}+
\frac{\delta S[L]}{\delta l_{ijk}}\delta l_{ijk}+\cdots
\end{equation}
and, since the fields $l_{i_1...i_n}$ can be varied independently, the equations
\begin{equation}
\frac{\delta S[L]}{\delta L}=0\hskip5pt\text{and}\hskip5pt
\Biggl\{ \frac{\delta S[L]}{\delta l_i}=0,\hskip5pt
\frac{\delta S[L]}{\delta l_{ij}}=0,\hskip5pt
\frac{\delta S[L]}{\delta l_{ijk}}=0,\hskip5pt ...\Biggr\}
\end{equation}
are equivalent. The variation of (\ref{S}) thus returns the tensorial field equations for each of the fields $l_{i_1...i_n}$. 

In the perturbative approach considered in this paper, the action (\ref{S}) can always be reduced to a standard spacetime action. In fact, given the series expansions (\ref{H}) and (\ref{bg}) for $H$ and $\bg_{ij}$ respectively, we can rewrite the Gibbs factor as
\begin{equation}
e^{-\kappa H}=e^{-\frac{1}{2}\kappa g_{ij}\dot{x}^i\dot{x}^j}\left(1-\frac{1}{3}\kappa l_{ijk}\dot{x}^i\dot{x}^j\dot{x}^k+\cdots\right)
\end{equation}
and the determinant of $\bg_{ij}$ as
\begin{equation}
\bg=g\left[1+{l^h}_{hi}\dot{x}^i+\left({l^h}_{hi}{l^k}_{kj}-\frac{1}{2}{l^h}_{ki}{l^k}_{hj}\right)\dot{x}^i\dot{x}^j+\cdots\right],
\end{equation}
with $g=\det g_{ij}$.
The integration along the fibers of $TM$ reduces then to the evaluation of pseudo-Gaussian integrals of the form $\int\dot{x}^{i_1}...\ \dot{x}^{i_k}e^{-\frac{1}{2}\kappa g_{ij}\dot{x}^i\dot{x}^j}d^4\dot{x}$,
that cancel out for $k\equiv 2n-1$ odd,
\begin{equation}
\label{pseudo-Gaussian-odd}
\int\dot{x}^{i_1}...\ \dot{x}^{i_{2n-1}}e^{-\frac{1}{2}\kappa g_{ij}\dot{x}^i\dot{x}^j}d^4\dot{x}=0,
\end{equation}
 and are all proportional to a single divergent  integral for $k\equiv 2n$ even
\begin{equation}
\label{pseudo-Gaussian-even}
\int\dot{x}^{i_1}...\dot{x}^{i_{2n}}e^{-\frac{1}{2}\kappa g_{ij}\dot{x}^i\dot{x}^j}d^4\dot{x}=\frac{(2n)!}{2^nn!\kappa^{n}\sqrt{
| g| }}g^{(i_1i_2}...\ g^{i_{2n-1}i_{2n})}
\int e^{-\frac{1}{2}\kappa\eta_{ab}y^ay^b}d^4y,
\end{equation}
with $\eta_{ab}=\text{diag}(1,-1,-1,-1)$ and round brackets again indicating symmetrization.
By reabsorbing the infinite constant factor in the normalization constant, i.e.\ by setting $\mathcal{N}(\kappa)=\int e^{-\frac{1}{2}\kappa\eta_{ab}y^ay^b}d^4y$, after the integration in the $\dot{x}$, the action (\ref{S}) reduces to the familiar form of an integral over the spacetime of an effective Lagrangian density for the geometric fields $l_{i_1...i_n}(x)$  and the matter fields under consideration.

\subsection{Einstein-Maxwell dynamics}

We are now in a position to demonstrate that it is indeed possible to promote a regular autonomous free Lagrangian to a dynamic field in such a way that the interactions of electromagnetic and gravitational type induced in the low kinetic energy-type regime evolve according to Einstein--Maxwell dynamics. In the low kinetic  energy-type limit the dynamics associated to any free Lagrangian is in fact described by the Lagrangian (\ref{L2}), $\bg_{ij}$ correspond to $g_{ij}$, $\bg=g$, the Hamiltonian reduces to $H=\frac{1}{2}g_{ij}\dot{x}^i\dot{x}^j$ and the curvature scalar $\br$ is given in  (\ref{bR2}). Ignoring for the moment the presence of matter, the substitution of $\br$, $H$, $\bg$ in (\ref{S}) yields the action
\begin{equation}
\label{Sqa}
S=-\frac{1}{\mathcal{N}(\kappa)}\iint_{TM} 
\biggl[
-\nabla_j{F^j}_i\dot{x}^i+\Bigl( R_{ij}+\frac{1}{2H}{F^h}_iF_{hj}\Bigr)\dot{x}^i\dot{x}^j
\biggr] e^{- \frac{1}{2}\kappa g_{kl}\dot{x}^k\dot{x}^l}g d^4xd^4\dot{x}.
\end{equation}
The integration in the $\dot{x}$ reduces to 
\begin{equation}
\label{integral1}
\int\dot{x}^ie^{- \frac{1}{2}\kappa g_{kl}\dot{x}^k\dot{x}^l}d^4\dot{x}=0,
\end{equation}
\begin{equation}
\label{integral2}
\int\dot{x}^i\dot{x}^je^{- \frac{1}{2}\kappa g_{kl}\dot{x}^k\dot{x}^l}d^4\dot{x}=\frac{1}{\kappa\sqrt{| g| }}g^{ij}\mathcal{N}(\kappa)
\end{equation}
and
\begin{equation}
\label{integral3}
\int\frac{\dot{x}^i\dot{x}^j}{g_{kl}\dot{x}^k\dot{x}^l}e^{- \frac{1}{2}\kappa g_{kl}\dot{x}^k\dot{x}^l}d^4\dot{x}=\frac{1}{4\sqrt{| g| }}g^{ij}\mathcal{N}(\kappa),
\end{equation}
where the first two integrals are given in (\ref{pseudo-Gaussian-odd}), (\ref{pseudo-Gaussian-even}) and the third one is formally reduced to the second one in the~\ref{App}. The tangent bundle action (\ref{Sqa}) reduces then to the spacetime action
\begin{equation}
\label{S2}
S=-\int_M\left(\frac{1}{\kappa}R+\frac{1}{4}F^{ij}F_{ij}\right)\sqrt{| g| }d^4x,
\end{equation}
that describes the classical Einstein--Maxwell dynamics for the gravitational and electromagnetic fields.  
It is possible to verify by direct calculation that the variation with respect to $l_i$ and $g_{ij}$ and the integration over the fibers commute and yield the Einstein--Maxwell equations
in both cases. The introduction of matter, so as the interpretation of an
$\dot{x}$ dependent matter Lagrangian density~\cite{LaemmerzahlandPerlick2018}, do not present difficulties. 
The promotion of a general free Lagrangian to a dynamical field provides therefore a coherent unified geometrical description of the gravitational and electromagnetic interactions at the classical level.

\section{Conclusions}

Using the free motion of elementary test particles in a four-dimensional spacetime as a guideline, we have shown that for small values of the energy-type first integral, identified with half of the square of the invariant mass of the particles in Planck units,
the paths in configuration space of any  Lagrangian system that satisfies the regularity conditions (i) and (ii)  and is analytic in the fiber coordinates $\dot{x}$, are indistinguishable from the geodesics of a mass-dependent Randers metric and therefore from the trajectories of a representative point subjected to background electromagnetic and gravitational interactions. 
Furthermore, we have shown that it is possible to promote the Lagrangian function, identified with the coefficients of its series  expansion, to a dynamic field, in such a way that these interactions evolve according to Einstein--Maxwell dynamics. 

Higher energy-type values introduce higher rank tensorial interactions that generalize the electromagnetic and the gravitational ones. In spacetime physics, however, given the small values of the  expansion parameter, the intensity of these interactions is several order of magnitude weaker than the gravitational one. On the one hand, this justifies the fact that these interactions, should they exist, have never been observed and therefore makes it possible that this mechanism is indeed at the origin of real electromagnetic and gravitational fields. On the other hand, it makes it difficult to decide whether a character of reality can be attributed to the higher rank interactions or whether these are of any practical use in the physics of spacetime. If they are, they describe Finsler corrections to classical fundamental interactions. To answer the question it is clearly necessary to carefully study their dynamics and evaluate their effects on the motion of particles and light, in cases where they may be relevant.
 Of equal importance to access the non-perturbative regime of the theory, would be a formulation directly in terms of the Lagrangian function and not of its expansion coefficients as done in this paper. Whatever the outcome will be, since the theory has no free parameters that are not fixed by the small $\mu$ regime, if the effects of the higher rank fields are sufficiently strong to be observed, they will immediately prove or disprove the validity of this approach. In any case, the general analysis we presented remain true for  all regular Lagrangian systems.

\appendix{}
\section[pfx={Appendix},nonum]{}
\label{App}

To formally evaluate (\ref{integral3}) we proceed as follows. For $H>0$ we rewrite $\frac{1}{H}e^{-\kappa H}=\int^{\infty}_\kappa e^{-z H}dz$,
\setcounter{equation}{0}
\numberwithin{equation}{section}
\begin{equation}
\int\frac{1}{2H}\dot{x}^i\dot{x}^je^{-\kappa H}d^4\dot{x}=
\frac{1}{2}\int \dot{x}^i\dot{x}^j \left[\int^{\infty}_\kappa e^{-z H}dz\right] d^4\dot{x}
\end{equation}
we invert the orders of integration, we again apply (\ref{pseudo-Gaussian-even}) with $n=1$
\begin{equation}
=\frac{1}{2}
\int^{\infty}_\kappa \left[\int \dot{x}^i\dot{x}^j  e^{-\frac{1}{2}z g_{ij}\dot{x}^i\dot{x}^j}d^4\dot{x}\right] dz=\frac{1}{2}
\int^{\infty}_\kappa \frac{1}{z\sqrt{| g| }}g^{ij}\mathcal{N}(z) dz
\end{equation}
and by observing that $\mathcal{N}(\kappa)=\frac{1}{\kappa^2}\mathcal{N}(1)$ we obtain
\begin{equation}
=
\frac{1}{2\sqrt{| g| }}g^{ij}\mathcal{N}(1) \int^{\infty}_\kappa\frac{1}{z^3}dz
=
\frac{1}{4\sqrt{| g| }}g^{ij}\frac{1}{\kappa^2}\mathcal{N}(1)
=
\frac{1}{4\sqrt{| g| }}g^{ij}\mathcal{N}(\kappa).
\end{equation}
For $H<0$ we set instead $\frac{1}{H}e^{-\kappa H}=-\int_{-\infty}^\kappa e^{-z H}dz$ and proceed analogously obtaining the same result.

\newpage

\end{document}